\documentclass[reprint,superscriptaddress,aps]{revtex4-1}
\usepackage{amsmath}
\usepackage{amssymb}
\usepackage{bm}
\usepackage{color}
\usepackage[varg]{txfonts}
\usepackage{varwidth}
\usepackage{dcolumn}
\usepackage[breaklinks,colorlinks=true,linkcolor=blue,urlcolor=cyan,citecolor=blue]{hyperref}
\usepackage[dvipdfmx]{graphicx}

\begin{document}

\title{Magnetization process of the breathing pyrochlore magnet CuInCr$_{4}$S$_{8}$\\
in ultra-high magnetic fields up to 150~T}

\author{Masaki Gen}
 \email{gen@issp.u-tokyo.ac.jp}
 \affiliation{Institute for Solid State Physics, University of Tokyo, Kashiwa, Chiba 277-8581, Japan}

\author{Yoshihiko Okamoto}
 \affiliation{Department of Applied Physics, Nagoya University, Nagoya 464-8603, Japan}
 
\author{Masaki Mori}
 \affiliation{Department of Applied Physics, Nagoya University, Nagoya 464-8603, Japan}
 
\author{Koshi Takenaka}
 \affiliation{Department of Applied Physics, Nagoya University, Nagoya 464-8603, Japan}

\author{Yoshimitsu Kohama}
 \email{ykohama@issp.u-tokyo.ac.jp}
 \affiliation{Institute for Solid State Physics, University of Tokyo, Kashiwa, Chiba 277-8581, Japan}

\date{\today}

\begin{abstract}

The magnetization process of the breathing pyrochlore magnet CuInCr$_{4}$S$_{8}$ has been investigated in ultra-high magnetic fields up to 150~T. 
Successive phase transitions characterized with a substantially wide 1/2-plateau from 55~T to 110~T are observed in this system, resembling those reported in chromium spinel oxides.
In addition to the 1/2-plateau phase, the magnetization is found to exhibit two inherent behaviors: a slight change in the slope of the $M$--$H$ curve at $\sim 85$~T and a shoulder-like shape at $\sim 130$~T prior to the saturation.
Both of them are accompanied by a hysteresis, suggesting first-order transitions.
The theoretical calculation applicable to CuInCr$_{4}$S$_{8}$ is also shown, based on the microscopic model with the spin-lattice coupling.
The calculation fairly well reproduces the main features of the experimentally observed magnetization process, including a relatively wide cant 2:1:1 phase clearly observed in the previous work [Y. Okamoto {\it et al.}, J. Phys. Soc. Jpn. {\bf 87}, 034709 (2018)].
The robust 1/2-plateau on CuInCr$_{4}$S$_{8}$ seems to be originated from the dominant antiferromagnetic interactions and the strong spin-lattice coupling.

\end{abstract}


\maketitle

\section{\label{sec:level1}INTRODUCTION}

Frustrated spin systems have been extensively studied for several decades because they can exhibit macroscopically degenerate ground states such as quantum spin-liquid \cite{1973_And, 1998_Moe}.
In real compounds, however, the macroscopic degeneracy is lifted by various perturbations such as quantum and thermal fluctuations \cite{1998_Moe}, spin-lattice coupling \cite{2002_Tch}, Dzyaloshinskii-Moriya interaction \cite{2005_Elh}, and so on.
Intriguingly, such perturbations can induce successive phase transitions under magnetic fields, including unconventional magnetic phases as represented by a magnetization plateau.

Chromium spinel oxides {\it A}Cr$_{2}$O$_{4}$ ({\it A} = Hg, Cd, Zn, Mg) are well known as typical 3D frustrated magnets exhibiting field-induced successive phase transitions \cite{2006_Ueda, 2011_Kim, 2014_Nak, 2008_Koj, 2013_Miy, 2011_Miy_JPSJ, 2011_Miy_PRL, 2012_Miy, 2014_Miy}.
In these systems, non-magnetic divalent cations occupying the tetrahedral {\it A} sites form a diamond lattice, whereas magnetic Cr$^{3+}$ ions octahedrally surrounded by oxygen ions form a pyrochlore lattice.
The orbital degrees of freedom are quenched because of the half-filled $t_{2g}$ orbitals, making Cr spinel oxides an ideal $S=3/2$ Heisenberg spin system with strong geometrical frustration.
This frustration suppresses a long range antiferromagnetic (AFM) ordering well below the Weiss temperature, i.e. $T_{\mathrm{N}}\ll |\Theta_{\mathrm{CW}}|$ \cite{2006_Ueda, 2005_Chu, 2000_Lee, 2009_Ji, 2008_Ort}.
The strength of the nearest-neighbor (NN) exchange interaction strongly depends on the Cr-Cr distance \cite{2008_Yar, 2008_Ueda}, indicative of a strong spin-lattice coupling, and consequently, the frustration is resolved due to the spin Jahn-Teller effect at $T_{\mathrm N}$,\cite{2008_Tan, 2007_Che}.
The spin-lattice coupling is also responsible for the robust 1/2-plateau phase with a 3 up-1 down spin configuration that appears universally in these oxide compounds \cite{2006_Ueda, 2011_Kim, 2014_Nak, 2008_Koj, 2013_Miy, 2011_Miy_JPSJ, 2011_Miy_PRL, 2012_Miy, 2014_Miy, 2004_Penc, 2006_Mot, 2006_Berg, 2007_Penc, 2010_Sha}.
Recently, {\it A}-site ordered Cr spinel oxides Li{\it M}Cr$_{4}$O$_{8}$ ({\it M} = In, Ga) have been attracted attention \cite{1966_Jou, 2013_Oka, 2014_Tan, 2015_Nil, 2016_Sah, 2016_Lee, 2017_Oka, 2019_Gen}.
In these systems, Cr$^{3+}$ ions form a breathing pyrochlore lattice comprised of an alternating array of small and large tetrahedra, where two kinds of NN AFM exchange interactions, $J$ and $J'$, exist, respectively \cite{2013_Oka}.
The difference in the strengths of $J$ and $J'$ might introduce unconventional magnetic properties in high magnetic fields, distinct from {\it A}Cr$_{2}$O$_{4}$.
However, since Li{\it M}Cr$_{4}$O$_{8}$ possesses extremely strong NN AFM interactions, as indicated by the large Weiss temperature ($\Theta_{\mathrm{CW}}=-332$~K and $-659$~K for {\it M} = In and Ga, respectively \cite{2013_Oka}), the observation up to saturation requires several hundred tesla, which has not yet been achieved experimentally \cite{private1}.

Here, we report a combined experimental and theoretical investigation into the magnetization process of an {\it A}-site ordered Cr spinel sulfide, CuInCr$_{4}$S$_{8}$, on which the Weiss temperature is known to be $-7(2) \times 10^{1}$~K \cite{2018_Oka}, suggesting experimentally accessible saturation field.
The fundamental magnetic properties of this compound were investigated in 1970s \cite{1970_Pin, 1971_Plu, 1975_Ung, 1977_Plu, 1980_Plu}, and refocused by Okamoto {\it et al.} recently \cite{2018_Oka}.
The heat capacity of CuInCr$_{4}$S$_{8}$ shows a sharp peak at $T_{\mathrm{p}}=28$~K with a sudden drop in the magnetic susceptibility, suggesting the magnetic ordering below $T_{\mathrm{p}}$ \cite{2018_Oka}.
The previous neutron diffraction experiment clarified that the spin structure at 4.2~K is a collinear order comprising decoupled (100) ferromagnetic planes \cite{1971_Plu, 1977_Plu}.
This strongly indicates the coexistence of AFM $J$ and FM $J'$ for small and large tetrahedra, respectively.
The magnetization process of CuInCr$_{4}$S$_{8}$ was revealed up to 73~T, where $M$ reaches 45~\% of the saturation magnetization of $M_{\mathrm{s}}=3.06~\mu_{\mathrm{B}}$/Cr (this value is estimated from the Land\'{e} $g$ factor of 2.04) \cite{2018_Oka}.
The magnetization shows a jump at $\sim25$~T with a large hysteresis loop and exhibits a kink at $\sim40$~T suggesting a first order phase transition.
Another magnetization kink appears at $\sim55$~T followed by a plateau-like behavior, which may be a phase transition to the 1/2-plateau state.
In this study, we have performed magnetization measurements on CuInCr$_{4}$S$_{8}$ under ultra-high magnetic fields up to 150~T to elucidate the full magnetization process.
We also theoretically investigated the magnetization process of the breathing pyrochlore magnet by using a microscopic model with the spin-lattice coupling.
Achievement of these works can help us to understand the high-field properties of a novel type of the pyrochlore spin system that consists of two kinds of NN exchange interactions with opposite sign.

This paper is organized as follows.
In section II, we show the experimental method and the result of high-field magnetization measurements.
The details of measurement technique are described in Supplemental Material.
In section III, we introduce the microscopic Heisenberg model incorporating the spin-lattice coupling, which can be applied to the breathing pyrochlore magnet with $J>0$ and $J'<0$.
The numerical calculation on the effective spin model gives a detailed and general phase diagram and magnetization curves for CuInCr$_{4}$S$_{8}$.
In section IV, we compare the calculated magnetization curves with those obtained in experiments and discuss the characteristic features for CuInCr$_{4}$S$_{8}$.
Finally, the possible spin structures on CuInCr$_{4}$S$_{8}$ under magnetic fields are proposed, and the strengths of several exchange interactions of CuInCr$_{4}$S$_{8}$ are also estimated.

\section{\label{sec:level2}EXPERIMENT}

A polycrystalline powder sample of CuInCr$_{4}$S$_{8}$, synthesized by a solid-state reaction method as in Ref. \cite{2018_Oka}, was used in the present work. 
The lattice parameter was found to be $a=10.05970(11)$~$\AA$, and the amount of intersite defects was estimated at most $\sim 3$~\%.
High-field magnetization measurements were performed using a horizontal single-turn-coil (HSTC) system up to 150~T.
The pulsed-field duration time was approximately 7.3~$\mu$s.
The induction method was adopted to detect the $dM/dt$ signal using a coaxial-type self-compensated magnetization pickup coil.
In order to minimize the uncompensated contribution of the background signal, three sets of measurements, in the order of sample-out, sample-in, and sample-out, were carried out as in Ref. \cite{2019_Gen}.
The detailed setup to obtain high-quality magnetization data under magnetic fields up to 150~T is described in Supplemental Material.
The magnetic field was measured by a calibrated pickup coil wound around the magnetization pickup coil.
The sample was cooled down to approximately 5~K using a liquid-He flow cryostat made of glass-epoxy (G-10).
The temperature was monitored by a RuO$_{2}$ resistance thermometer.

\begin{figure}[t]
 \centering
 \hfill
 \includegraphics[width=0.9\columnwidth]{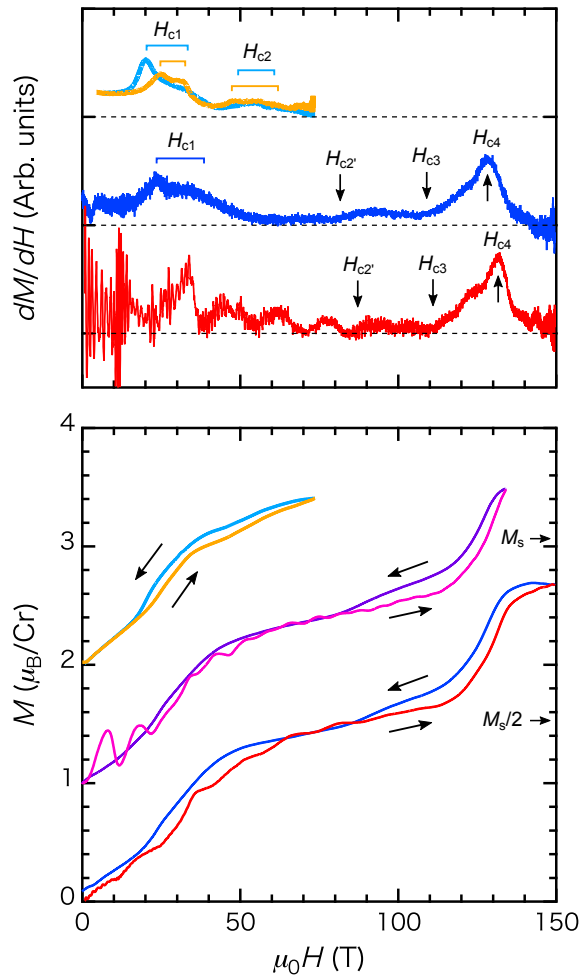}
 \hfill\null
 \caption{$M$--$H$ curves of a powder sample of CuInCr$_{4}$S$_{8}$ measured at approximately 5~K in a HSTC megagauss generator. Measurements were performed up to 134~T (pink and purple curves for up and down sweeps, respectively) and 150~T (red and blue curves for up and down sweeps, respectively). The $dM/dH$ of the $M$--$H$ curves up to 150~T are shown in the upper panel. The previously reported $M$--$H$ curve and its derivative $dM/dH$ measured at 1.4~K in a non-destructive pulsed magnet up to 73~T are also shown (orange and cyan curves for up and down sweeps, respectively) \cite{2018_Oka}. Each data is shifted vertically for clarity (The dashed lines in the upper panel indicate $dM/dH=0$ lines). Transition fields are denoted by brackets or arrows in $dM/dH$.}
\label{CuInCr4S8}
\end{figure}

The $M$--$H$ curves (up to 134~T and 150~T) and the derivative $dM/dH$ (up to 150~T) of CuInCr$_{4}$S$_{8}$ are shown in Fig. \ref{CuInCr4S8}.
They are consistent with the previous result up to 73~T obtained by a non-destructive pulsed magnet \cite{2018_Oka}, which is also shown in Fig. \ref{CuInCr4S8}.
Although the present data taken in HSTC have poor Signal-to-Noise ratio up to $\sim 60$~T for up sweep, the two sets of data overlap quantitatively in the high field region up to 134~T, guaranteeing the high accuracy of our measurements.
As seen in Fig. \ref{CuInCr4S8}, the hysteresis opening at $\sim 20$~T once closes in the field region of $60\sim80$~T, where $M$ is almost constant at $\sim M_{\mathrm{s}}/2$, implying the 1/2-plateau phase. 
For both up and down sweeps, this plateau-like feature survives until $\sim 110$~T, then $M$ clearly exhibits an upturn behavior.
Note that a substantial hysteretic behavior is seen in the middle of the plateau region: a gradual $dM/dH$ change is seen at $\sim 87$~T for up sweep and at $\sim 81$~T for down sweep.
With increasing a magnetic field above the plateau phase, $M$ exhibits a shoulder shape at $\sim 130$~T, followed by a gradual increase until 150~T, where $M$ reaches $\sim 2.7~\mu_{\mathrm{B}}$/Cr.
Assuming that $M$ increases linearly above 150~T, the saturation field is expected to be $\sim 180$~T. 
Judging from the fact that $M$ reaches the value of $M_{\mathrm{s}}/2$ at $\sim 90$~T for up sweep, this estimation could be plausible.

\begin{table}[t]
 \caption{Transition fields for CuInCr$_{4}$S$_{8}$ determined from the anomalies in $dM/dH$. The unit is tesla.}
  \begin{tabular}{|c|c|c|c|c|c|c|} \hline
     & $H_{\mathrm{c1}}$ & $H_{\mathrm{c2}}$ & $H_{\mathrm{c2'}}$ & $H_{\mathrm{c3}}$ & $H_{\mathrm{c4}}$ & $H_{\mathrm{sat}}$ \\ \hline
    up & ~29$\pm$4~ & ~55$\pm$7~ & ~87$\pm$4~ & ~111$\pm$2~ & ~132 $\pm$2~ & ~$>$150~ \\
    down & ~27$\pm$7~ & ~55$\pm$6~ & ~81$\pm$2~ & ~109$\pm$2~ & ~128 $\pm$2~ & ~$>$150~ \\ \hline
    order & 1st & 2nd & 1st & 2nd & 1st & --- \\ \hline
  \end{tabular}
 \label{tab:transition fields}
\end{table}

We summarize the transition fields for CuInCr$_{4}$S$_{8}$ in Table \ref{tab:transition fields}, which are experimentally determined from the anomalies in $dM/dH$.
$H_{\mathrm{c1}}$ and $H_{\mathrm{c2}}$ for both up and down sweeps are deduced from the previous result in Ref. \cite{2018_Oka} by taking the value around the hump in $dM/dH$ (the upper panel in Fig. \ref{CuInCr4S8}).
The order of each transition is judged from the existence of a hysteresis.
Note that our magnetization data did not detect the hump structure around $H_{\mathrm{c2}}$ for down sweep.
This might be the influence of the fast field-sweep rate or the magnetic field inhomogeneity which becomes inevitable for the down sweep due to the deformation of the field generation coil.
Interestingly, there are in total six phase transitions up to the full saturation, resulting in a very complicated magnetization process.
The detailed explanation on the experimental result and its theoretical analysis are given in the following sections.

\section{\label{sec:level3}CALCULATION}

As mentioned above, chromium spinel oxides  {\it A}Cr$_{2}$O$_{4}$ exhibit a variety of field-induced magnetic phases, represented by a robust 1/2-plateau phase.
Although several theoretical studies have been devoted to the Cr spinel oxides \cite{2004_Penc, 2006_Mot, 2006_Berg, 2007_Penc, 2010_Sha, 2015_Tak}, the ground state of the breathing pyrochlore under magnetic fields has not been investigated so far.
For the sake of the interpretation of the observed magnetization process, we constructed an effective spin model on a breathing pyrochlore magnet  with $J>0$ and $J'<0$ and examined the ground state under magnetic fields.

\subsection{BP model and SP model}

First, we introduce two microscopic models considering the spin-lattice coupling used for a standard pyrochlore antiferromagnet.
One was first proposed by Penc {\it et al.} \cite{2004_Penc}, called the bond-phonon (BP) model, which assumes independent changes in the distance between neighboring spins ${\mathbf S}_{i}$ and ${\mathbf S}_{j}$.
In the following, we treat spins in the classical limit, and normalize to $|{\mathbf S}|=1$.
The effective Heisenberg Hamiltonian of the BP model is expressed as 
\begin{equation}
\label{eq:H_BP}
{\mathcal{H}}_{\mathrm{BP}}=J\sum_{\langle i, j\rangle}[{\mathbf S}_{i} \cdot {\mathbf S}_{j}-b({\mathbf S}_{i} \cdot {\mathbf S}_{j})^2]-{\mathbf h}\sum_{i}{\mathbf S}_{i},
\end{equation}
where the summation ${\langle i, j\rangle}$ is taken over all the NN bonds, the coefficient $b$ of the biquadratic term is a dimensionless parameter representing the strength of the spin-lattice coupling
\begin{equation}
\label{eq:parameter_b}
b=\frac{1}{cJ} \left[\left.\frac{dJ}{dr}\right|_{r=|{\mathbf r}_{ij}^{0}|}\right]^{2},
\end{equation}
where $c$ is an elastic constant, $|{\mathbf r}_{ij}|^{0}$ is the bond length between NN sites at their regular positions.
When we take $J>0$ and ${dJ}/{dr}<0$, $b$ becomes a positive value.
In applied magnetic fields, the BP model can determine the local spin structure on each tetrahedron.
As magnetic field is increased, the model undergoes successive phase transitions from an AFM phase to cant 2:2, 1/2-plateau, cant 3:1, then spin-saturated phases.
In a weak spin-lattice coupling regime ($b\lesssim 0.05$), a cant 2:1:1 phase appears between the cant 2:2 and the 1/2-plateau phase.
The mechanism for stabilizing the 1/2-plateau phase can be attributed to the biquadratic term favoring a collinear spin configuration.
The zero-temperature $b$--$h$ phase diagram on Eq. \ref{eq:H_BP} has been well understood \cite{2004_Penc, 2007_Penc, 2010_Sha}, and the corresponding magnetization curves are compatible to the experimentally observed magnetization processes of {\it A}Cr$_{2}$O$_{4}$ ($b=$0.15, 0.10, and 0.02 for {\it A} = Hg, Cd, and Zn, respectively).
However, the biquadratic term cannot completely reproduce the degeneracy lifting in the pyrochlore system because spin correlations beyond NN sites via the lattice distortion are not taken into account on the BP model, resulting in the absence of the magnetic long-range order \cite{2010_Sha}.
This is not the case in real compounds.

Alternatively, Bergman {\it et al.} \cite{2006_Berg} proposed another microscopic spin model, called the site-phonon (SP) model, which assumes independent displacement of each site position.
The effective Hamiltonian of the SP model is expressed as
\begin{equation}
\begin{split}
\label{eq:H_SP}
{\mathcal{H}}_{\mathrm{SP}}&=J\sum_{\langle i, j\rangle}[{\mathbf S}_{i} \cdot {\mathbf S}_{j}-b({\mathbf S}_{i} \cdot {\mathbf S}_{j})^2]\\
&\quad-J\frac{b}{2}\sum_{j\neq k\in N(i)}{\mathbf e}_{ij} \cdot {\mathbf e}_{ik}({\mathbf S}_{i} \cdot {\mathbf S}_{j})({\mathbf S}_{i} \cdot {\mathbf S}_{k})\\
&\quad-{\mathbf h}\sum_{i} {\mathbf S}_{i},
\end{split}
\end{equation}
where ${\mathbf e}_{ij}$ denotes the unit vector connecting NN sites $i$ and $j$ at their regular positions, $N(i)$ denotes the set of NN sites of site $i$.
In addition to the biquadratic term as present in the BP model, the SP model includes an additional 3-body term derived from the effective second and third NN interactions due to the lattice distortion.
Since the biquadratic term still plays a dominant role on the SP model, the basic feature of the magnetization curves remains unchanged.
On the other hand, the additional term in Eq. \ref{eq:H_SP} reduces the macroscopic degeneracy of the spin degrees of freedom, leading to the magnetic long-range order.
For example, as a long-range AFM state at zero field, a tetragonal collinear spin structure with (1,1,0) magnetic Bragg peaks is predicted for $b<0.25$ \cite{2016_Aoyama}.
This is consistent with the neutron scattering experiments on {\it A}Cr$_{2}$O$_{4}$ where the observed Bragg-peak patterns involve (1,1,0) reflections although they are composed of rich and complex reflections \cite{2005_Chu, 2007_Mat, 2008_Lee}.
Furthermore, the SP model predicts a 16-sublattice cubic spin structure with space group $P4_{3}32$ for the 1/2-plateau phase.
This spin structure was also observed in high-field neutron scattering experiments on HgCr$_{2}$O$_{4}$ and CdCr$_{2}$O$_{4}$ \cite{2007_Mat, 2010_Mat}.

\subsection{Effective spin model applicable to CuInCr$_{4}$S$_{8}$}

Here, let's move to the case of the breathing pyrochlore lattice.
Recently, Aoyama {\it et al.} \cite{2019_Aoyama} derived a SP model in the presence of breathing lattice distortion.
The Heisenberg Hamiltonian considering two kinds of NN exchange interactions, $J$ and $J'$, is written as
\begin{equation}
\begin{split}
\label{eq:H_0}
{\mathcal{H}_{0}}=J\sum_{\langle i, j\rangle_{\mathrm{S}}}{\mathbf S}_{i} \cdot {\mathbf S}_{j}+J'\sum_{\langle i, j\rangle_{\mathrm{L}}}{\mathbf S}_{i} \cdot {\mathbf S}_{j},
\end{split}
\end{equation}
where the summation ${\langle i, j \rangle_{\mathrm{S}}}$ (${\langle i, j \rangle_{\mathrm{L}}}$) is defined only in the small (large) tetrahedra.
Assuming $J, J'>0$ and ${dJ}/{dr}, {dJ'}/{dr}<0$, we can define two spin-lattice coupling parameters with positive sign, $b$ and $b'$, in the small and large tetrahedra, respectively.
Then, the spin interactions mediated by the SP effect can be expressed as
\begin{equation}
\begin{split}
\label{eq:H_SLC}
{\mathcal{H}_{\mathrm{SLC}}}&=-Jb\sum_{\langle i, j\rangle_{\mathrm{S}}}({\mathbf S}_{i} \cdot {\mathbf S}_{j})^2-J'b'\sum_{\langle i, j\rangle_{\mathrm{L}}}({\mathbf S}_{i} \cdot {\mathbf S}_{j})^2\\
&\quad-\sum_{i} \left\{ \frac{Jb}{4}\sum_{j\neq k\in N_{\mathrm{S}}(i)}+\frac{J'b'}{4}\sum_{j\neq k\in N_{\mathrm{L}}(i)} \right\} ({\mathbf S}_{i} \cdot {\mathbf S}_{j})({\mathbf S}_{i}\cdot {\mathbf S}_{k})\\
&\quad-\sqrt{JJ'bb'}\sum_{i}\sum_{j\in N_{\mathrm{S}}(i)}\sum_{k\in N_{\mathrm{L}}(i)}{\mathbf e}_{ij} \cdot {\mathbf e}_{ik}({\mathbf S}_{i} \cdot {\mathbf S}_{j})({\mathbf S}_{i} \cdot {\mathbf S}_{k}),\\
\end{split}
\end{equation}
where $N_{\mathrm{S}}(i)$ ($N_{\mathrm{L}}(i)$) is defined only in the small (large) tetrahedra.
The derivation process of Eq. \ref{eq:H_SLC} is described in Ref. \cite{2019_Aoyama}.
As well as the case of {\it A}Cr$_{2}$O$_{4}$, this Hamiltonian also stabilizes a tetragonal collinear spin structure with (1,1,0) magnetic Bragg peaks in the wide ranges of $b$, $b'$, and $J'/J$ \cite{2019_Aoyama}, which is in agreement with the experimentally observed domain state in the low-temperature ordered phase on Li{\it M}Cr$_{4}$O$_{8}$ ({\it M}=In, Ga) \cite {2015_Nil, 2016_Sah}.
The investigation of $b$--$h$ phase diagrams on the SP model for various values of $J'/J$ ($0<J'/J\leq 1$) is in progress \cite{private2}. 

Although one adopted this model to the antiferromagnet with $J, J'>0$, it is also applicable to the case of $J>0$ and $J'<0$.
In this process, however, we have to be careful of the preceding sign of each term in Eq. \ref{eq:H_SLC}.
In the case of $J'<0$, the sign of ${dJ'}/{dr}$ is nontrivial for the following reasons.
In Cr spinel sulfides, the NN exchange interaction is mainly originated from the AFM direct exchange interaction between NN Cr sites and the FM superexchange interaction via Cr-S-Cr path.
The difference in the Cr-Cr distance on the small and large tetrahedra is only 6~\% at room temperature on CrInCr$_{4}$S$_{8}$, so the opposite sign of $J$ and $J'$ implies that the AFM direct exchange and the FM superexchange interactions are competitive.
Considering that the former is affected by the Cr-Cr distance whereas the latter by the Cr-S-Cr angle, which is clarified to be more than $90^{\circ}$ in all NN Cr pairs on  CuInCr$_{4}$S$_{8}$ \cite {2018_Oka}, the increase in the Cr-Cr distance will make both the direct exchange and the superexchange interactions weaker.
Hence, both situations, ${dJ'}/{dr}<0$ and ${dJ'}/{dr}>0$, could be realized in CrInCr$_{4}$S$_{8}$.
Note that the above explanation cannot remove the possibility of ${dJ}/{dr}>0$, but the assumption of ${dJ}/{dr}<0$ is more plausible because the direct exchange interaction becomes relatively dominant as the Cr atoms get closer.
Regardless of the sign of ${dJ}/{dr}$ and ${dJ'}/{dr}$, the sign of each spin-lattice coupling parameter becomes $b>0$ and $b'<0$ (See Eq. \ref{eq:parameter_b}).
However, under the assumption of ${dJ}/{dr}<0$ and ${dJ'}/{dr}>0$, the preceding sign of the last term in Eq. \ref{eq:H_SLC} changes form minus to plus \cite{proof}.
In the following discussion, we will exclude this case, i.e. we will assume ${dJ}/{dr}<0$ and ${dJ'}/{dr}<0$ as the effect of the spin-lattice coupling.
Indeed, the calculation for the case of ${dJ'}/{dr}<0$ reproduces the experimental results better than the case of ${dJ'}/{dr}>0$, in the sense that it can reproduce the observed wide intermediate phase prior to the 1/2-plateau phase on CuInCr$_{4}$S$_{8}$.

\begin{figure}[t]
 \centering
 \hfill
 \includegraphics[width=0.9\columnwidth]{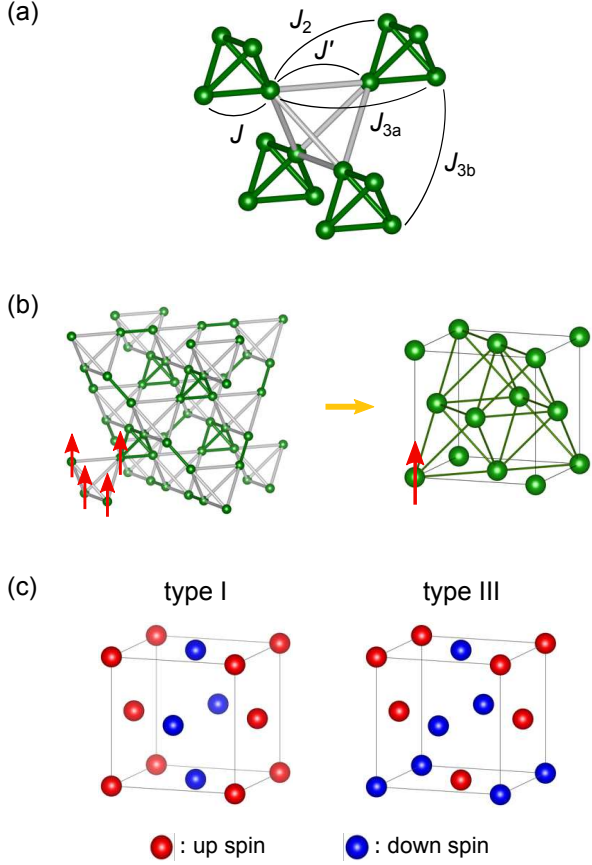}
 \hfill\null
 \caption{(a) Crystal structure of the breathing pyrochlore lattice with two kinds of NN exchange interactions, $J$ and $J'$, in the small and large tetrahedra, respectively. Second NN ($J_{2}$) and third NN interactions ($J_{3a}$ and $J_{3b}$ for two symmetrically inequivalent paths) are also shown. (b) Model transformation from the breathing pyrochlore magnet with $J>0$ and $J'<0$ to the fcc lattice antiferromagnet. Four spins in each large tetrahedron are converted to one localized spin, as indicated by red arrows. (c) Two types of AFM ordering which are candidate for the ground state on the fcc lattice antiferromagnet. Sites with up (down) spins are colored by red (blue).}
\label{lattice}
\end{figure}

In addition, we include further-neighbor (FN) AFM interactions between sites of second and third NNs, which are expected to be strong in sulfides unlike oxides \cite{2008_Yar, 2019_Gho}:
\begin{equation}
\label{eq:H_FN}
{\mathcal{H}_{\mathrm{FN}}}=J_{2}\sum_{\langle\langle i, j\rangle\rangle}{\mathbf S}_{i} \cdot {\mathbf S}_{j}+J_{3a}\sum_{{\langle\langle\langle i, j\rangle\rangle\rangle}_{3a}}{\mathbf S}_{i} \cdot {\mathbf S}_{j}+J_{3b}\sum_{{\langle\langle\langle i, j\rangle\rangle\rangle}_{3b}}{\mathbf S}_{i} \cdot {\mathbf S}_{j},
\end{equation}
where the summations ${\langle\langle i, j\rangle\rangle}$, ${\langle\langle\langle i, j\rangle\rangle\rangle}_{3a}$, and ${\langle\langle\langle i, j\rangle\rangle\rangle}_{3b}$ are taken for second NN and two kinds of third NN sites ($3a$ and $3b$), respectively [Fig. \ref{lattice}(a)].
For simplicity, we do not take care of the effects of the magnetostriction on the strengths of these FN interactions.
Organizing the above, we obtain
\begin{equation}
\label{eq:H_eff1}
{\mathcal{H}_{\mathrm{CuInCr_{4}S_{8}}}}={\mathcal{H}_{0}}+{\mathcal{H}_{\mathrm{SLC}}}+{\mathcal{H}_{\mathrm{FN}}}-{\mathbf h}\sum_{i}{\mathbf S}_{i}
\end{equation}
as the spin Hamiltonian of CuInCr$_{4}$S$_{8}$ under magnetic fields.

In this study, we convert this complicated model into a simple one.
We naively anticipate that all four spins in the same large tetrahedron always take a ferromagnetically aligned spin configuration even in the external magnetic fields as in zero field.
By treating those four spins as one localized spin at the center of a large tetrahedra, the breathing pyrochlore magnet with $J>0$ and $J'<0$ can be mapped onto the antiferromagnet composed of a face-centered cubic (fcc) lattice [Fig. \ref{lattice}(b)].
In the following, we will represent a spin vector ${\mathbf S}_{\alpha}$ as a spin located at the fcc site after the model transformation mentioned above.
By omitting constant terms and assuming $Jb=J'b'$ in Eq. \ref{eq:H_eff1}, we finally derive the effective spin Hamiltonian of CuInCr$_{4}$S$_{8}$ as
\begin{equation}
\begin{split}
\label{eq:H_eff2}
{\mathcal{H}_{\mathrm{CuInCr_{4}S_{8}}}^{\mathrm{eff}}}&=(J+4J_{2}+2J_{3a}+2J_{3b})\sum_{\langle \alpha, \beta\rangle}{\mathbf S}_{\alpha} \cdot {\mathbf S}_{\beta}\\
&\quad-Jb\sum_{\langle \alpha,\beta\rangle}({\mathbf S}_{\alpha} \cdot {\mathbf S}_{\beta})^2+4Jb\sum_{\langle \alpha, \beta\rangle}{\mathbf S}_{\alpha} \cdot {\mathbf S}_{\beta}\\
&\quad-\frac{Jb}{4}\sum_{k}\sum_{\alpha\neq\beta\neq \gamma\in k}({\mathbf S}_{\alpha} \cdot {\mathbf S}_{\beta})({\mathbf S}_{\alpha}\cdot {\mathbf S}_{\gamma})\\
&\quad-\frac{\mathbf h}{4}\sum_{\alpha}{\mathbf S}_{\alpha},
\end{split}
\end{equation}
where the first summation in the fourth term are taken for all local tetrahedra $k$ depicted in the fcc lattice of Fig. \ref{lattice}(b).
The second term is originated from the BP effect, and the third and fourth terms from the SP effect.
Note that the fcc lattice can be regarded as the 3D network of edge-sharing tetrahedra, possessing geometrical frustration.
If we consider only NN interactions on the fcc lattice, two kinds of Neel orders, Type I and Type III [Fig. \ref{lattice}(c)], become candidate for the ground state at zero field (Type II state can only be realized in the existence of FN interactions).
Such a degeneracy can be lifted by fluctuations \cite{1998_Moe, 1987_Hen, 2015_Ben}, FN interactions \cite{1980_Plu}, and so on.
Theoretically, It has been shown that fluctuations favor Type I state with an ordering wavevector {\bf q}$=(1,0,0)$.
Indeed, the observed magnetic structure of CuInCr$_{4}$S$_{8}$ at zero field can be mapped onto Type I state \cite{1977_Plu}.

\subsection{Calculation results}

In this section, we focus on the characteristics of the zero-temperature $b$--$h$ phase diagram and magnetization curves derived from Eq. \ref{eq:H_eff2}.
In our analysis, the fourth and sixth NN interactions in the original breathing pyrochlore lattice are not taken into account.
Consequently, only NN interactions and intra-tetrahedral interactions appear in the effective Hamiltonian on the fcc lattice.
Thus, we can rewrite Eq. \ref{eq:H_eff2} by taking a summation of local Hamiltonian on each single tetrahedron:
\begin{equation}
\label{eq:H_eff3}
{\mathcal{H}_{\mathrm{CuInCr_{4}S_{8}}}^{\mathrm{eff}}}=\sum_{k}{\mathcal{H}_{\mathrm{CuInCr_{4}S_{8}}}^{\mathrm{local}}},
\end{equation}

\begin{equation}
\begin{split}
\label{eq:H_local}
{\mathcal{H}_{\mathrm{CuInCr_{4}S_{8}}}^{\mathrm{local}}}&=[J(1+4b)+J_{\mathrm{FN}}]\sum_{\langle \alpha, \beta\rangle_{k}}{\mathbf S}_{\alpha} \cdot {\mathbf S}_{\beta}\\
&\quad-Jb\sum_{\langle \alpha,\beta\rangle_{k}}({\mathbf S}_{\alpha} \cdot {\mathbf S}_{\beta})^2\\
&\quad-\frac{Jb}{4}\sum_{\alpha\neq\beta\neq \gamma\in k}({\mathbf S}_{\alpha} \cdot {\mathbf S}_{\beta})({\mathbf S}_{\alpha}\cdot {\mathbf S}_{\gamma})\\
&\quad-\frac{\mathbf h}{4}\sum_{\alpha\in k}{\mathbf S}_{\alpha},
\end{split}
\end{equation}
where $J_{\mathrm{FN}}\equiv 4J_{2}+2J_{3a}+2J_{3b}$, and the summation $\langle \alpha, \beta\rangle_{k}$ is taken over all pairs in a single tetrahedron $k$. 
Here, we normalize spin vectors to $|{\mathbf S}_{\alpha}|=$1, and choose $J_{\mathrm{FN}}/J$ and $b$ as adjustable dimensionless parameters.
Since all the optimum spin configuration on each local tetrahedron can be simultaneously satisfied on the fcc lattice with an infinite size, we can obtain the ground state just by numerically minimizing Eq. \ref{eq:H_local} in a certain magnetic field.

\begin{figure}[t]
 \centering
 \hfill
 \includegraphics[width=0.8\columnwidth]{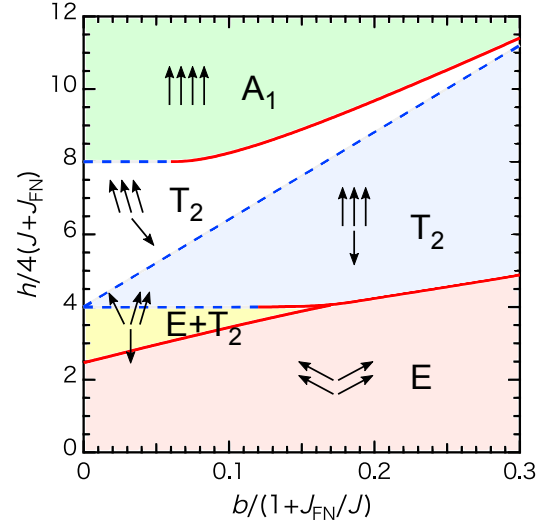}
 \hfill\null
 \caption{Phase diagram of Eq. \ref{eq:H_local} as a function of the spin-lattice coupling parameter $b$ and magnetic field $h$. Red solid and blue dashed lines denote first- and second-order transitions, respectively. The regions shaded in pink, yellow, blue, white, and green express a cant 2:2, cant 2:1:1, 1/2-plateau, cant 3:1, and fully polarized phase, respectively. Schematic spin configuration within a single tetrahedron and its irreducible representation of the tetrahedral symmetry group are also illustrated in each region. This phase diagram is applicable to any value of $J_{\mathrm{FN}}$.}
\label{PD}
\end{figure}

\begin{figure}[t]
 \centering
 \hfill
 \includegraphics[width=0.8\columnwidth]{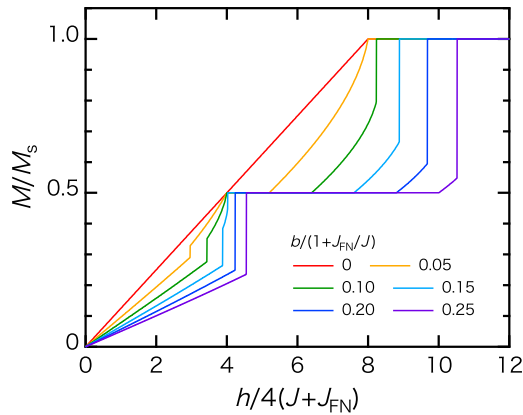}
 \hfill\null
 \caption{Magnetization curves as a function of magnetic field $h$ for $b/(1+J_{\mathrm{FN}}/J)=0$ to 0.25 in steps of 0.05.}
\label{MHcurve}
\end{figure}

The $b$--$h$ phase diagram and the corresponding magnetization curves for given values of $b$ are summarized in Figs. \ref{PD} and \ref{MHcurve}, respectively.
Regardless of the value of $J_{\mathrm{FN}}/J$, the $b$--$h$ phase diagram becomes identical except the scale of both axes.
Here, we demonstrate the case of $J_{\mathrm{FN}}/J=0$.
As shown in Fig. \ref{PD}, several magnetic phases such as cant 2:2, cant 2:1:1, 1/2-plateau, cant 3:1, and spin-saturated phases appear in applied magnetic fields.
We define $h_{\mathrm{c1}}$, $h_{\mathrm{c2}}$, $h_{\mathrm{c3}}$, and $h_{\mathrm{sat}}$ as transition fields to the cant 2:1:1, 1/2-plateau, cant 3:1, and spin-saturated phases, respectively. 
For $b\lesssim 0.17$, the cant 2:1:1 phase appears immediately below the 1/2-plateau phase.
The transition from the cant 2:2 to cant 2:1:1 phase at $h_{\mathrm{c1}}$ is the first order accompanied by a magnetization jump.
The transition from the cant 2:1:1 to 1/2-plateau phase at $h_{\mathrm{c2}}$ is the second order for $b\lesssim 0.12$, where the value of $h_{\mathrm{c2}}/4$ is constant to $4J$, while it turns to the first order for $b\gtrsim 0.12$, where $h_{\mathrm{c2}}$ becomes slightly higher as $b$ increases.
The transition field $h_{\mathrm{c1}}$ gets monotonously higher as $b$ increases, and finally marges with $h_{\mathrm{c2}}$ at $b\approx 0.17$.
For $b\gtrsim 0.17$, a first-order transition from the cant 2:2 to 1/2-plateau phase occurs at $h_{\mathrm{c2}}$, accompanied by a magnetization jump.
The width of the 1/2-plateau phase is extremely broad, indicating that the 3 up-1 down collinear spin configuration with the magnetization $M/M_{\mathrm{s}}=1/2$ is substantially stable in a wide field region.
When higher magnetic field is applied in the 1/2-plateau phase, the system undergoes a second-order transition to the cant 3:1 phase at $h_{\mathrm{c3}}/4=4(1+6b)J$, then finally enters the spin-saturated phase at $h_{\mathrm{sat}}$.
The transition from the cant 3:1 to spin-saturated phase is the second order for $b\lesssim 0.06$, where the value of $h_{\mathrm{sat}}/4$ is constant to $8J$, while it becomes the first order for $b\gtrsim 0.06$, where $h_{\mathrm{sat}}$ increases gradually as $b$ increases.

Although the $b$--$h$ phase diagram of Eq. \ref{eq:H_local} is similar with that of Eq. \ref{eq:H_BP} proposed in Ref. \cite{2004_Penc}, there are two remarkable differences. 
First, as $b$ increases the broadening of the 1/2-plateau phase becomes saturated for Eq. \ref{eq:H_BP}, whereas the width of the 1/2-plateau becomes constantly wider in our model.
Second, the cant 2:1:1 phase emerges in wider ranges of $b$ and $h$ in our results.
If we adopt the BP model to the breathing pyrochlore magnet with $J>0$ and $J'<0$, the effective Hamiltonian becomes identical with Eq. \ref{eq:H_BP}.
Hence, the differences in the $b$--$h$ phase diagram can be considered to originate from the effective FN interactions due to the lattice distortion, which is only taken into account on the SP model (The third and fourth terms of Eq. \ref{eq:H_eff2}).

\section{\label{sec:level4}DISCUSSION AND CONCLUSION}

A few experimental studies have been made under high magnetic fields on CuInCr$_{4}$S$_{8}$ so far \cite{1980_Plu, 2018_Oka}.
Despite these efforts, there is still little understanding on the field-induced magnetic phases.
Here, we compare our experimental and theoretical results on CuInCr$_{4}$S$_{8}$ with those on Cr spinel oxides, and discuss how our research contributes to the fundamental understanding of the magnetization process of CuInCr$_{4}$S$_{8}$.

In the low field region up to 70~T, two phase transitions at $H_{\mathrm{c1}}$ and $H_{\mathrm{c2}}$ are obviously seen prior to the 1/2-plateau.
The former is the first-order phase transition which is accompanied by a large hysteresis and associated with a cusp in $dMdH$, whereas the latter might be the second order one with a broad hump in $dM/dH$ (visible only in the result obtained by a non-destructive pulsed magnet \cite{2018_Oka}).
The calculated magnetization curve with a small spin-lattice coupling parameter ($b\lesssim 0.17$) shown in Fig. \ref{MHcurve} well reproduces this behavior qualitatively.
Hence, the intermediate phase between $H_{\mathrm{c1}}$ and $H_{\mathrm{c2}}$ can be attributed to a cant 2:1:1 phase.
The cant 2:1:1 phase has already been observed for ZnCr$_{2}$O$_{4}$ ($120\sim 135$~T) \cite{2011_Miy_JPSJ} and MgCr$_{2}$O$_{4}$ ($125\sim 140$~T) \cite{2014_Miy} by optical Faraday rotation measurements, but its field region is narrow compared to the cant 2:2 phase.
In contrast, the $M$--$H$ curve of CuInCr$_{4}$S$_{8}$ seems to show a relatively broad cant 2:1:1 phase although it is difficult to define the precise values of $H_{\mathrm{c1}}$ and $H_{\mathrm{c2}}$.
Such a broad cant 2:1:1 phase is reproduced better by our theoretical calculation based on the SP model rather than the BP model \cite{2004_Penc}.
As shown in Fig. \ref{MHcurve}, if the magnetization changes linearly in higher fields than $h_{\mathrm{c1}}$, it will reach $M_{\mathrm{s}}/2$ at around $H_{\mathrm{c3}}$ regardless of the value of $b$.
On the other hand, the linear fit of the experimental $M$--$H$ curve below $H_{\mathrm{c1}}$ will cross the $M=M_{\mathrm{s}}/2$ value at $\sim 70$~T, which is much lower than the value of the observed $\mu_{0}H_{\mathrm{c3}} \approx 110$~T.
There are a few possible reasons for this behavior.
One is that we restrict our theoretical analysis on the simple case of the equal coupling in different symmetry $b_{\mathrm{A_{1}}}=b_{\mathrm{E}}=b_{\mathrm{T_{2}}}$.
When we adopt the microscopic model to real compounds, we can assume different values of $b$, i.e. $b_{\mathrm{A_{1}}}\neq b_{\mathrm{E}}\neq b_{\mathrm{T_{2}}}$ \cite{2004_Penc, 2007_Penc, 2015_Kim}.
Indeed, $b_{\mathrm{T_{2}}}$ seems larger than $b_{\mathrm{E}}$ for HgCr$_{2}$O$_{4}$ and CdCr$_{2}$O$_{4}$ \cite{2015_Kim}, which might also be true for CuInCr$_{4}$S$_{8}$.
Another possible reason is that AFM FN interactions might become stronger in the 1/2-plateau phase due to the lattice distortion.
In order to accurately account for the observed value of $H_{\mathrm{c3}}$. it is necessary to estimate the change in the FN interactions caused by the magnetostriction.

In addition to the phase transitions mentioned above, two exotic features are observed in the $M$--$H$ curve of CuInCr$_{4}$S$_{8}$, which cannot be explained by our theoretical calculation.
First, a gradual $dM/dH$ change appears at $H_{\mathrm{c2'}}$ accompanied by a substantial hysteresis.
Judging from the second-order-like behavior (non-hysteretic) at $H_{\mathrm{c2}}$ and $H_{\mathrm{c3}}$, it is natural to consider that $H_{\mathrm{c2}}$ and $H_{\mathrm{c3}}$ correspond to $h_{\mathrm{c2}}$ and $h_{\mathrm{c3}}$ defined in our calculated results, respectively.
Accordingly, the magnetic structure between $H_{\mathrm{c2}}$ and $H_{\mathrm{c3}}$ should be the 3 up-1 down spin configuration, analogous to the 1/2-plateau phase realized in Cr spinel oxides.
One possible mechanism for the first-order nature at $H_{\mathrm{c2'}}$ is that the global symmetry of the spin structure changes accompanied by the lattice distortion while maintaining the collinear 3:1 spin configuration in small tetrahedra.
Such a scenario can happen in certain special situations: for example, FM $J'$ turns to AFM at $H_{\mathrm{c2'}}$.
To the best of our knowledge, a phase transition within a magnetization plateau, in which the size of the magnetization remains unchanged, has been reported only on MnCr$_{2}$S$_{4}$, which is composed of two kinds of magnetic ions, Mn$^{2+}$ ($S=5/2$) and Cr$^{3+}$ ($S=3/2$) occupying the tetrahedral $A$ and octahedral $B$ sites of the spinel structure {\it A}{\it B}$_{2}${\it X}$_{4}$, respectively \cite{2017_Tsu}.
In the case of MnCr$_{2}$S$_{4}$, however, the transition within the magnetization plateau was hardly detected as anomalies in the $M$--$H$ curve while it was clearly observed by the ultrasound measurements. 
As for CuInCr$_{4}$S$_{8}$, a dedicated technique sensitive to the lattice deformation under high magnetic fields might be required to understand such an exotic feature in the $M$--$H$ curve.
Second, the $M$--$H$ curve of CuInCr$_{4}$S$_{8}$ exhibits a shoulder-like behavior characterized with a sharp peak in $dM/dH$ at $H_{\mathrm{c4}}$, where $M$ reaches $\sim 2.5~\mu_{\mathrm{B}}$/Cr, which is much smaller than $M_{\mathrm{s}}=3.06~\mu_{\mathrm{B}}$/Cr.
It suggests the existence of another intermediate phase between the cant 3:1 and spin-saturated phases.
In the series of Cr spinel oxides {\it A}Cr$_{2}$O$_{4}$ ({\it A} = Hg, Cd, Zn), a shoulder-like behavior in the $M$--$H$ curve and a drastic change in the intensity of the optical absorption associated with an exciton-magnon-phonon process have been reported just before the full saturation of magnetization \cite{2014_Nak, 2013_Miy, 2011_Miy_PRL}.
This intermediate phase is now believed to be a spin-nematic phase, after the theoretical proposal incorporating the quantum effect \cite{2015_Tak}.
Recently, the ground state of the $S=1/2$ Heisenberg model on the fcc lattice under magnetic fields was also investigated by Morita {\it et al.} \cite{2019_Mor}. 
The magnetization curve on the $S=1/2$ fcc lattice without tetragonal distortion is quite similar with our results (Fig. \ref{MHcurve}), and exhibits a cant 2:2, cant 2:1:1, 1/2-plateau, cant 3:1, and spin-saturated phase.
In this case, the tetragonal distortion on the fcc lattice induces more diverse super-solid phases immediately below the 1/2-plateau and the spin-saturated phase. 
This distortion can be regarded as the lattice distortion caused by the spin-lattice coupling on the isotropic fcc lattice.
Thus, these quantum phases might be relevant to our experimental observation although CuInCr$_{4}$S$_{8}$ is a $S=3/2$ system.

Finally, we remark on the estimation of the exchange interactions on CuInCr$_{4}$S$_{8}$.
Under the mean-field approximation, the Weiss temperature $\Theta_{\mathrm{CW}}$ and the saturation field $H_{\mathrm{sat}}$ can be deduced as follows, respectively:
\begin{equation}
\begin{split}
\label{eq:Weiss}
\Theta_{\mathrm{CW}}=-\frac{S(S+1)}{k_{\mathrm{B}}}(J+J'+J_{\mathrm{FN}}),
\end{split}
\end{equation}

\begin{equation}
\label{eq:H_sat}
\mu_{0}H_{\mathrm{sat}}=\frac{8Sk_{\mathrm{B}}}{g\mu_{\mathrm{B}}}(J+J_{\mathrm{FN}}),
\end{equation}
where $k_{\mathrm{B}}$ is the Boltzmann's constant and $\mu_{\mathrm{B}}$ is the Bohr magneton.
By combining two formulas Eqs. \ref{eq:Weiss} and \ref{eq:H_sat}, $J+J_{\mathrm{FN}}$ and $J'$ can be obtained independently.
Previously, Plumier {\it et al.} \cite{1980_Plu} calculated them from the experimental results of the magnetic susceptibility and the magnetization up to 38~T by using the estimated values of $\Theta_{\mathrm{CW}}=-77$~K and $\mu_{0}H_{\mathrm{sat}}=146$~T, where they might mistakenly multiply a factor of 2 to the right side of Eq. \ref{eq:Weiss}.
Here, we recalculate in the same way, which yields the exchange interactions of $J+J_{\mathrm{FN}}=21$~K and $J'= -2$~K with $g=2.04$, $\Theta_{\mathrm{CW}}=-7 \times 10^{1}$~K \cite{2018_Oka}, and $\mu_{0}H_{\mathrm{sat}}=1.8 \times 10^{2}$~T (estimated from this study).
This implies that the AFM interactions are dominant on CuInCr$_{4}$S$_{8}$, which could be responsible for a robust 1/2-plateau as observed in Cr spinel oxides.
It should be noted that the $M$--$H$ curves of LiGaCr$_{4}$S$_{8}$ and LiInCr$_{4}$S$_{8}$, where the FM interaction $J'$ is expected to be strong, do not clearly exhibit the 1/2-plateau \cite{2018_Oka}.
Recently, the exchange interactions of several Cr spinel compounds forming a breathing pyrochlore lattice were theoretically investigated by Ghosh {\it et al.} \cite{2019_Gho}.
By using the lattice parameter of CuInCr$_{4}$S$_{8}$ at room temperature, the exchange interactions are obtained as $J=14.7$~K, $J'=-26.0$~K, $J_{2}=1.1$~K, $J_{3a}=6.4$~K and $J_{3b}=4.5$~K.
Assuming these values, the saturation field is calculated to be $\sim 3.6 \times 10^{2}$~T, which is much higher than the experimental observation.
It seems that we have to carefully consider the influence of the spin-lattice coupling and thermal fluctuation.

To summarize, we have investigated the magnetization process of the breathing pyrochlore magnet CuInCr$_{4}$S$_{8}$ with $J>0$ and $J'<0$ from both experimentally and theoretically.
The observed $M$--$H$ curve is characterized with a wide 1/2-plateau exhibiting from $\mu_{0}H_{\mathrm{c2}}\approx 55$~T to $\mu_{0}H_{\mathrm{c3}}\approx 110$~T, and the saturation field is estimated to be $\mu_{0}H_{\mathrm{sat}}\approx 180$~T.
Two unique behaviors are also observed in the $M$--$H$ curve: a slight slope change accompanied by a hysteresis at $\mu_{0}H_{\mathrm{c2'}}\approx 85$~T in the 1/2-plateau region, and a shoulder-like shape at $\mu_{0}H_{\mathrm{c4}}\approx 130$~T prior to the saturation.
In particular, there are few reports on the former phenomenon, i.e. a phase transition within a magnetization-plateau state.
We also have proposed the microscopic classical spin model to understand the magnetic behavior of the breathing pyrochlore magnet with $J>0$ and $J'<0$.
The calculation well reproduces the main features of the magnetization process of CuInCr$_{4}$S$_{8}$, especially a relatively wide cant 2:1:1 phase.
However, there are still many unsolved issues in CuInCr$_{4}$S$_{8}$, such as the global spin structure in each phase, the possible change in the exchange interactions under magnetic fields, and so on.
Further experimental investigation such as the magnetostriction, magnetocaloric effect, ultrasound, ESR, NMR, neutron and X-ray measurements under high magnetic fields should be interesting and will provide useful clues to understand the essence of the successive phase transitions on CuInCr$_{4}$S$_{8}$.

\section*{ACKNOWLEDGMENTS}

We appreciate for fruitful discussion to Dr. K. Morita, Dr. K. Aoyama, and Prof. H. Kawamura.
This work was supported by JSPS KAKENHI Grant Number 16H03848, 18H01163, and 19H05823.

\end{document}


\title{Supplemental Material for "Magnetization process of the breathing pyrochlore magnet CuInCr$_{4}$S$_{8}$\\
in ultra-high magnetic fields up to 150~T"}

\author{Masaki Gen}
 \email{gen@issp.u-tokyo.ac.jp}
 \affiliation{Institute for Solid State Physics, University of Tokyo, Kashiwa, Chiba 277-8581, Japan}

\author{Yoshihiko Okamoto}
 \affiliation{Department of Applied Physics, Nagoya University, Nagoya 464-8603, Japan}
 
\author{Masaki Mori}
 \affiliation{Department of Applied Physics, Nagoya University, Nagoya 464-8603, Japan}
 
\author{Koshi Takenaka}
 \affiliation{Department of Applied Physics, Nagoya University, Nagoya 464-8603, Japan}

\author{Yoshimitsu Kohama}
 \email{ykohama@issp.u-tokyo.ac.jp}
 \affiliation{Institute for Solid State Physics, University of Tokyo, Kashiwa, Chiba 277-8581, Japan}

\date{\today}

\begin{abstract}

\end{abstract}

\maketitle
\section{Improvement of magnetization measurement technique}

Magnetization is one of the most fundamental physical quantities in magnetic materials, and its precise observation is very important.
However, the direct magnetization measurement using a pickup coil under ultra-high magnetic field above 100 T is still challenging.
The difficulty comes from (i) large initial electro-magnetic noise in a single-turn-coil (STC) system, (ii) large induced voltage generated in the pickup coil due to the instantaneous magnetic field sweep, and (iii) the growth of the field inhomogeneity in space and time especially for down sweep due to the explosion of the STC.
Especially, the effect of (ii) and (iii) is enhanced as the generated magnetic field becomes higher.
In order to overcome these problems, we have already developed a coaxial-type self-compensated magnetization pickup coil instead of a parallel-type and realized precise measurements up to 130 T by using a STC with a diameter of 14 mm \cite{2019_Gen}.
The pickup coil was comprised of an inner-coil with 28 turns (diameter of 1.3 mm) and an outer-coil with 18 turns (diameter of 1.7 mm), using a polyamide-imide enameled copper wire (AIW wire, TOTOKU Electric Co. Ltd.) with an outer diameter of 0.06 mm and a reinforced insulation coating (withstand voltage of approximately 1 kV).
The deference in the winding numbers between inner- and outer-coils allows the detection of the magnetization signal from the sample.

In this work, we optimized the size and the winding number of the coaxial-type pickup coil to extend the measurable maximum field up to 150 T.
In order to measure above 130 T in STC system, we have to use a STC with a diameter of less than 14 mm (In this work, we adopt 12 mm).
Hence, we reduced the winding numbers of the pickup coil to prevent the dielectric breakdown of the wire insulation caused by high induced voltage, and changed the diameter of the outer coil from 1.7 mm to 2.1 mm to keep the deference in the winding numbers between inner- and outer-coils as much as possible.
The characteristics of the pickup coil used in this experiments are as follows: an inner-coil with 19 turns (diameter of 1.3 mm) and an outer-coil with 8 turns (diameter of 2.1 mm) for measurements up to 134 T, and an inner-coil with 14 turns (diameter of 1.3 mm) and an outer-coil with 6 turns (diameter of 2.1 mm) for measurements up to 150 T.
These pickup coils enable accurate magnetization measurements without the dielectric breakdown during several pulsed-field generation.